\documentclass[pre,amssymb,floats,twocolumn,showpacs,superscriptaddress]{revtex4}

\usepackage{epsfig}

\begin{document}

\title{Social inertia in collaboration networks}

\author{Jos{\'e} J. Ramasco}
\email{jose.ramasco@emory.edu}
\affiliation{Physics Department, Emory University, Atlanta Georgia 30322.}

\author{Steven A. Morris}
\affiliation{Oklahoma State University, Electrical and Computer Engineering,
Stillwater Oklahoma 74078.}

\date{\today}

\begin{abstract}

This work is a study of the properties of collaboration networks
employing the formalism of weighted graphs to represent their
one--mode projection. The weight of the edges is directly the number
of times that a partnership has been repeated. This representation
allows us to define the concept of {\it social inertia} that measures the
tendency of authors to keep on collaborating with previous
partners. We use a collection of empirical datasets to analyze
several aspects of the social inertia: 1) its
probability distribution, 2) its correlation with other properties, and
3) the correlations of the inertia between neighbors in the network.
We also contrast these empirical results with the predictions of a
recently proposed theoretical model for the growth of collaboration
networks.

\end{abstract}


\maketitle

\section{Introduction}

The study of complex networks has recently raisen a great
interest in a very multidisciplinary community (for recent reviews on the
field see \cite{barabasi02,sergei03,romu04,newman03}). Complex
network theory provides mathematical tools to directly deal with such
intricate systems as for instance the Internet or the World-Wide Web
\cite{barabasi99,albert99,romu01}. These two cases are real incarnations of
mathematical graphs, however the applicability of the theory
may be extended to many other situations. Actually, any system composed
of a set of interacting elements may be represented as a graph. The vertices 
correspond to the basic objects
in the system, and the edges model the interactions among them.
Protein interaction networks constitute a good example of how
insight may be gained into the micro and macro behavior of a
massively complex system using graph theory
\cite{jeong00,schwikowski00,wuchty03}.

Human society is also an extraordinarily complex system that can be
analyzed using the same theoretical concepts. In this particular case, the
vertices represent individuals and the edges social interactions
such as professional, friendship, or family relationships. The study
of such networks promises to provide quantitative understanding of
human collective behavior. To date, the biggest problem of studying
social systems has been the absence of large databases from which
reliable statistical conclusions could be drafted. Nevertheless, for
a special sort of social networks, the so called 
{\it collaboration networks}, that restriction no longer exists . The
current size of the digital databases is big enough to allow the
statistical characterization of the network topology and the
comparison of empirical results with the predictions produced by
theoretical models.

Collaboration networks are composed by two kind of vertices: 1)
{\it actors}, which are the persons involved in collaborations
(such as movie or theater actors, paper or book authors, football
players, or corporate board members), and 2) the 
{\it collaboration acts} (movies, theater performances, scientific 
papers, books,
common membership in a football team, or common membership on a
corporate board). In a collaboration network, the undirected edges
connect the actors to the collaborations in which they have taken
part. The fact that there exist two very different type of vertices
in the network is a central property that determines its structure
\cite{newman01,newman03b}. For this reason, these networks are
usually known as bipartite graphs and they are just a very
particular class of a wider set of complex networks with a variety
of vertex types. In order to study their topology, the standard
method is to perform a one--mode projection of the original network
where only the nodes representing actors remain and are connected to
each other whenever they share the same collaboration
\cite{barabasi99,newman01b}. Since this procedure neglects multiple
common collaborations, the resulting projected graph is less
informative than the original bipartite network. A way to partially avoid such
a loss of information is the use a weighted network for the
projection \cite{note0}. While collaboration graphs were originally studied as
having binary weighted links: 0 for no collaborations, 1 for one or
more collaborations, weighted networks are graphs in which each link
is associated with an {\it edge weight}, whose magnitude  can range
from 0 to infinity \cite{yook01,barrat04}.  In  this work, the link
weight in the projected network is the number of times a
collaboration between two actors has been repeated. In collections
of journal papers, it represents the number of papers that a pair of
authors have published together. Hence, the study of the weight 
distribution and
how it relates to the number of different co-actors allows to
extract information about the level of conservatism of the people at the
hour of collaborate with different partners, a property that we will call 
{\it social inertia}.

This paper is organized as follows: in Sec. \ref{weight} we
introduce the concepts and magnitudes that we are going to use
to analyze one-mode projected networks as weighted graphs. In the
Section \ref{data}, we present the results of analysis of a
collection of empirical collaboration networks. After that, in Sec.
\ref{models}, the results obtained from a theoretical model are
compared with those coming out from the empirical study. Finally, in
Sec. \ref{conclusion}, we end by discussing the significance of
social inertia as a model and empirical phenomenon.

\section{One-mode projections of collaboration networks as weighted graphs}
\label{weight}

Let us discuss first some of the quantities used to characterize
collaboration networks. A typical network is composed of $N_c$
collaborations and of $N_a$ actors. Of these actors, not all have
collaborators, $N_{ai}$ of them work alone. There are several degree 
distribution functions characterizing the network, two of which are 
fundamental. The first, 
$P_n(n)$, describes the probability that a collaboration has a given size 
$n$. While the 
second, $P_q(q)$, represents the probability that an actor has
participated in a total of $q$ collaborations. As a result of the 
one-mode projection, another
degree distribution may be defined for the projected network 
$P_k(k)$. Note that the meaning of $k$ is the number of different partners
that an actor (author) has had during his/her carrier.

\begin{table*}
\label{dat-tab} 
\caption{Global parameters of our set of empirical
databases.}
\begin{ruledtabular}
\begin{tabular}{ccccccccccccccc}
Field& $N_c$& $N_a$& $N_{ai}/N_{a} (\%)$& $m$& $\langle n \rangle$&
$\langle q \rangle$& $\langle k \rangle$& $R_{ks}$& $\langle
\mathcal{R} \rangle$& $\langle \mathcal{I} \rangle$& $\delta$&
$\langle C \rangle$& $\langle C^w \rangle$&
Ref.\\
\hline &
\multicolumn{14}{l}{{\it IMDB movie database}} \\

movies& $127823$& $383640$& $0.37$ & $3.0$& $11.5$& $3.83$& $78.4$&
$0.908$& $0.02$& $1.027$& $4.0$& $0.78$& $0.39$&
\cite{barabasi99,note1}\\

&
\multicolumn{14}{l}{{\it Scientific collaborations}} \\

anthrax& $2460$& $4320$& $8.9$ & $1.76$& $3.07$& $1.75$& $5.59$&
$0.62$& $0.08$& $1.16$& $3.9$& $0.79$& $0.40$&
\cite{nets-Stev}\\

atrial ablation& $3091$& $6409$& $0.78$ & $2.07$& $5.43$& $2.62$&
$9.18$& $0.48$& $0.14$& $1.33$& $3.0$& $0.84$& $0.43$&
\cite{nets-Stev,atrial-dis}\\

biosensors& $5889$& $10993$& $1.1$ & $1.87$& $3.89$& $2.08$& $6.05$&
$0.74$& $0.11$& $1.22$& $3.6$& $0.83$& $0.42$&
\cite{nets-Stev}\\

botox& $1560$& $3521$& $2.3$ & $2.26$& $3.84$& $1.7$& $5.74$&
$0.80$& $0.075$& $1.14$& $3.4$& $0.85$& $0.43$&
\cite{nets-Stev,botox}\\

complex networks& $900$& $1354$& $5.3$ & $1.51$ & $2.53$& $1.68$&
$3.15$& $0.052$& $0.089$& $1.19$& $3.52$& $0.69$& $0.35$&
\cite{nets-Stev,complex}\\

condmat& $22002$& $16721$& $2.8$ & $0.76$ & $2.66$& $3.50$& $5.69$&
$0.63$& $0.02$& $1.44$& $3.4$& $0.64$& $0.33$&
\cite{newman01b,note2}\\

distance education& $1389$& $2466$& $21.5$ & $1.78$ & $2.04$&
$1.15$& $2.56$& $0.96$& $0.02$& $1.04$& $\sim 6$& $0.66$& $0.33$&
\cite{nets-Stev,atrial-dis}\\

info science& $14209$& $9399$& $40.4$ & $0.66$ & $1.38$& $2.08$&
$1.58$& $0.85$& $0.06$& $1.12$& $3.9$& $0.48$& $0.24$&
\cite{nets-Stev}\\

info viz& $2448$& $5520$& $12.4$ & $2.26$ & $2.59$& $1.15$& $3.60$&
$0.94$& $0.02$& $1.04$& $\sim 5.2$& $0.77$& $0.39$&
\cite{nets-Stev}\\

scientometrics& $3467$& $2926$& $21.04$ & $0.84$ & $1.74$& $2.06$&
$2.20$& $0.78$& $0.08$& $1.18$& $3.5$& $0.54$& $0.28$&
\cite{nets-Stev}\\

self organized criticality& $1631$& $2040$& $5.4$ & $1.25$ & $2.57$&
$2.05$& $3.53$& $0.38$& $0.14$& $1.31$& $\sim 3.3$& $0.67$& $0.34$&
\cite{nets-Stev}\\

silicon on isolator& $2381$& $4867$& $1.3$ & $2.04$ & $4.0$& $1.95$&
$6.21$& $0.73$& $0.11$& $1.22$& $\sim 5$& $0.83$& $0.42$&
\cite{nets-Stev}\\

superconductors& $1629$& $2981$& $6.5$ & $1.83$ & $2.91$& $1.59$&
$4.88$& $0.81$& $0.08$& $1.16$& $4.1$& $0.80$& $0.40$&
\cite{nets-Stev}\\

superstrings& $6643$& $3755$& $7.8$ & $0.57$ & $2.04$& $3.62$&
$3.7$& $0.028$& $0.16$& $1.34$& $3.5$& $0.5$& $0.26$&
\cite{nets-Stev,superstring}\\

\end{tabular}
\end{ruledtabular}
\end{table*}

As explained above, we will consider the one-mode projected network as a
weighted graph. For an edge between
actor $i$ and actor $j$, the weight, $w_{ij}$, will be equal to the
number of collaborations between them.  Notice that this 
definition is different from the one widely used in the literature
\cite{newman01b,barthelemy04,fan04,borner05}.  Once the weight of links 
is defined, we
can also study its distribution, $P_w(w)$, as the probability that a
randomly chosen edge has a certain weight $w$. The existence of a
weight for the edges may change the importance of the vertices
within the network. Typically, the most significant nodes of a
graph, at least from a transport point of view, are those with the
highest number of connections, the hubs. However, in weighted
networks the link degree is not necessarily the most central
property in that sense. If the dispersion of $P_w(w)$ is very high,
it may be preferable to have {\it "high quality"} connections to your
neighbors, even if few in number, than having {\it "low quality"} 
connections to many neighbors.
To take this fact into account, another metric characterizing the
vertices is defined. This new variable is called {\it vertex strength} and 
measures a combination of weight and number of edges.
For a particular vertex $i$, the strength of $i$ is defined as:
\begin{equation}
s_i = \sum_{j \in{\mathcal V}(i)} w_{ij} ,
\end{equation}
where the sum runs over the set of all neighbors of $i$, 
${\mathcal V}(i)$. 

The strength of a vertex 
denotes the total number of partnerships (papers or movies) in which a 
particular actor
has been involved. This magnitude together with the degree of the nodes in the
one-mode projected network $k_i$, which contains the information
about how many of those partnerships have been with different
persons, allow us to define a measure of the conservatism of an
actor $i$ that
 we will call social inertia:
\begin{equation}
\mathcal{I}_i = s_i/k_i .
\end{equation}
The inertia is a new quantity that can be defined in general for all weighted
graphs but that has a very special meaning for the social networks. Its range
 goes from  $\mathcal{I}_i = 1$, in the
case of newcomers and actors that never repeat a collaborator, to
$q_i$, if all his/her collaborations were carried out always with
the same team. The higher $\mathcal{I}_i$ is, the more the actor
$i$ repeats his/her collaborators and consequently the more
conservative actor $i$ is about working with new people.
$\mathcal{I}_i$ is also related to the probability that the actor
$i$ repeats with one of his/her former collaborators by the
expression $\mathcal{R}_i = 1 - (k_i/s_i) = 1-(1/\mathcal{I}_i)$. It
is important to stress here that $1-\langle \mathcal{R} \rangle$ is
not equal the global ratio between the number of edges and the
number of partnerships in the networks expressed as
\begin{equation}
R_{ks} = \frac{\mbox{Total number of edges}}{\mbox{Total number of partnerships}} =
\frac{\langle k \rangle}{\langle s \rangle} .
\end{equation}
As happens with the previously introduced parameters, it
is possible to define probability distributions for finding a node
with a certain strength value $s$, $P_s(s)$ or a certain inertia
$\mathcal{I}$, $P_\mathcal{I}(\mathcal{I})$. However, only three of
the previous distributions are {\it a priori} independent (in the
absence of correlations among the different variables), let us say
$P_n(n)$, $P_q(q)$ and $P_w(w)$. The others should be derived from
these three basic functions.

In addition to the probability distributions, we also measure some other
quantities that further characterize the topology of networks. The clustering
is a good example of such magnitudes. The clustering is the density of
triangles in the network and hence it estimates how far the graph
is from a tree-like structure. For a vertex $i$, the definition of
the clustering of $i$ is given by
\begin{equation}
\label{clus}
c_i = \frac{2 \, t_i}{k_i \, (k_i-1)} ,
\end{equation}
where $t_i$ is the number of connections between the neighbors of $i$. This
concept may be generalized for weighted networks by means of the following
expression \cite{barrat04}
\begin{equation}
\label{weightc}
c_i^w = \frac{1}{s_i \, (k_i-1)} \sum_{j,m \in{\mathcal V}(i)} \frac{(w_{ij}+w_{im})}{2} \,
a_{ij} \, a_{im} \, a_{jm} ,
\end{equation}
where $a_{ij}$ is equal to one only if there is an edge between the
vertices $i$ and $j$ and zero otherwise. It is simple to check that
the definition of $c_i^w$ reduces to that of $c_i$ when there is
just one possible value for the weight of the links. These two
previous definitions refer only to local quantities, they can be
easily transformed into global parameters by averaging over
all the vertices of the network. In this way, we define 
the global clustering $C = \langle c_i \rangle$ and global weighted
clustering $C^w = \langle c_i^w \rangle$. For a good
comprehensive review on the characterization of complex weighted
networks, see \cite{barrat04}.

Other significant aspects of the network topology are the
correlations of the main properties like degree, strength or inertia
among neighboring vertices. The mean degree of the nearest neighbors
is a very informative magnitude in this respect
\cite{romu01,newman03b,newman02}. For a vertex $i$, it is defined as
\begin{equation}
k_{nn,i} = \frac{1}{k_i} \sum_{j \in{\mathcal V}(i)} k_j  .
\end{equation}
This quantity may also be expressed as a function of the degree,
$k_{nn}(k)$, by averaging over all nodes of the network with degree
$k$. If there are positive degree correlation between neighboring
nodes (the high degree vertices tend to be connected to high degree
vertices or {\it assortative mixing}), $k_{nn}(k)$ should grow
with increasing $k$. The contrary trend should be observed if the
network shows anticorrelation between the degree of the neighbors
({\it disassortative mixing}). The same general idea may be
applied to other properties of the vertices \cite{newman03c} as for
instance, the inertia of the nearest neighbors as a function of the own
inertia, $\mathcal{I}_{nn}(\mathcal{I})$, which is of special
relevance for this work.

\section{Empirical results}
\label{data}

\begin{figure}
\epsfxsize=80mm \epsffile{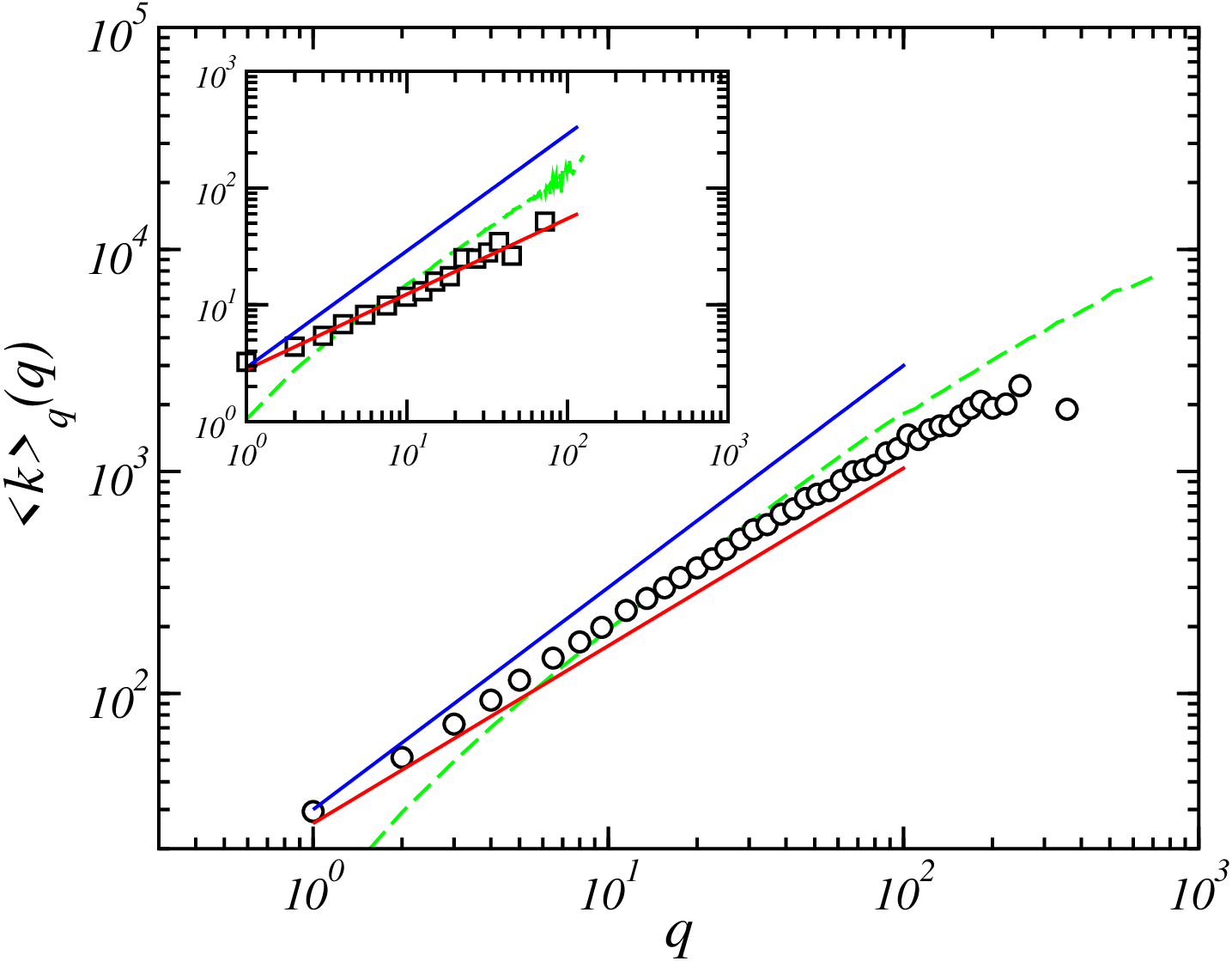}
\caption{Mean degree of the
actors in the one-mode projected network as a function of their
experience. In the main plot, for the movie database and in the
inset for the condmat database. In both cases, the blue (continuous) lines
correspond to a linear relation, the red (continuous) ones to power laws with
exponents $0.8$ in the main plot and $0.65$ in the inset and the
green (dashed) curves are the results from the model simulation.} \label{kq}
\end{figure}

In this section, we analyze some databases covering a range of
social communities. As may be seen in Table I, our biggest network
corresponds to the IMDB database on movies that includes as many as
$127823$ productions. In this case, the collaboration acts are
movies or TV series in which actors, previously hired by a producer,
perform. Here the decision mechanism about the cast is different from
the process of selecting authors for scientific collaborations. The
question about who is going to be an author of a scientific paper is
typically decided in a more self-organized way. This fact explains
the low average inertia that the movie network displays, $\langle
\mathcal{I} \rangle = 1.045$. However, it is important to note that 
the low value of $\langle \mathcal{I} \rangle$ is the only 
distinguishable characteristic of the movie
database over the others. In particular, a higher or lower value of
$\langle \mathcal{I} \rangle$ does not imply that the distribution
$P_\mathcal{I}(\mathcal{I})$ is short tailed. Another striking
aspect of Table I already among the scientific networks refers to
the disparity in the individualist ratio ($N_{ai}/N_s$) between
social and natural science papers, being much higher in the former
ones. Although a marked difference in that ratio (almost a factor
two) may be either observed between experimental and theoretical
works on natural sciences. Regarding the inertia among scientific
collaborations, authors on social sciences (those that do not 
work alone) tend to show the lowest
average values of $\mathcal{I}$ followed by the experimental
articles of natural sciences. For topics in the same area, 
$\mathcal{I}$ could be probably also 
related to the dynamism/age of the field.

\begin{figure}
\epsfxsize=80mm \epsffile{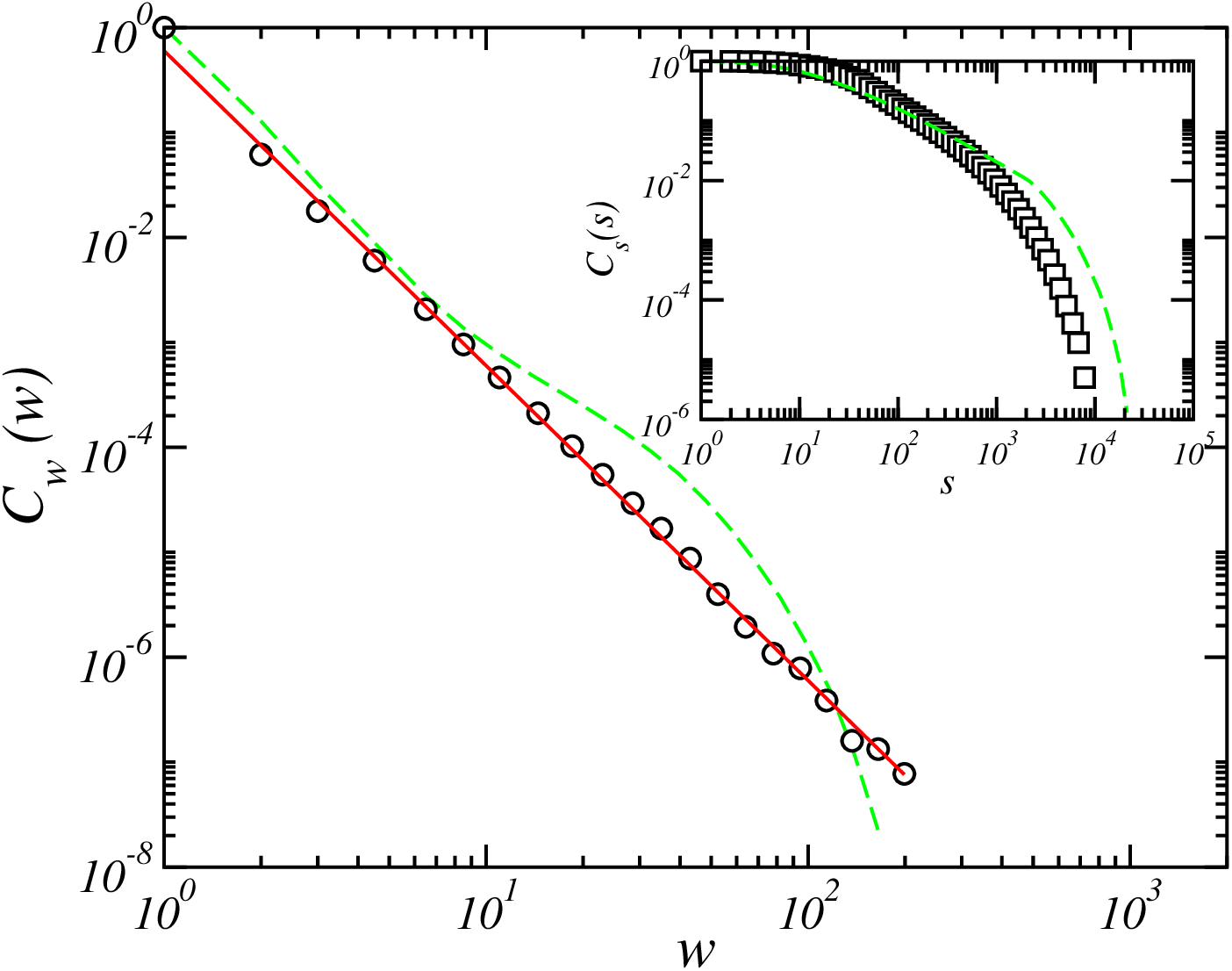}
\caption{Cumulative distribution for the weight of the edges in the movie
network. In the inset, the same distribution but for strength of the
nodes. The red (continuous) line has a slope of $1-\delta = -3$. The green
(dashed) curves are
always (for all figures) numerical results from the theoretical model. }
\label{pw}
\end{figure}

\begin{figure}[b!]
\epsfxsize=80mm \epsffile{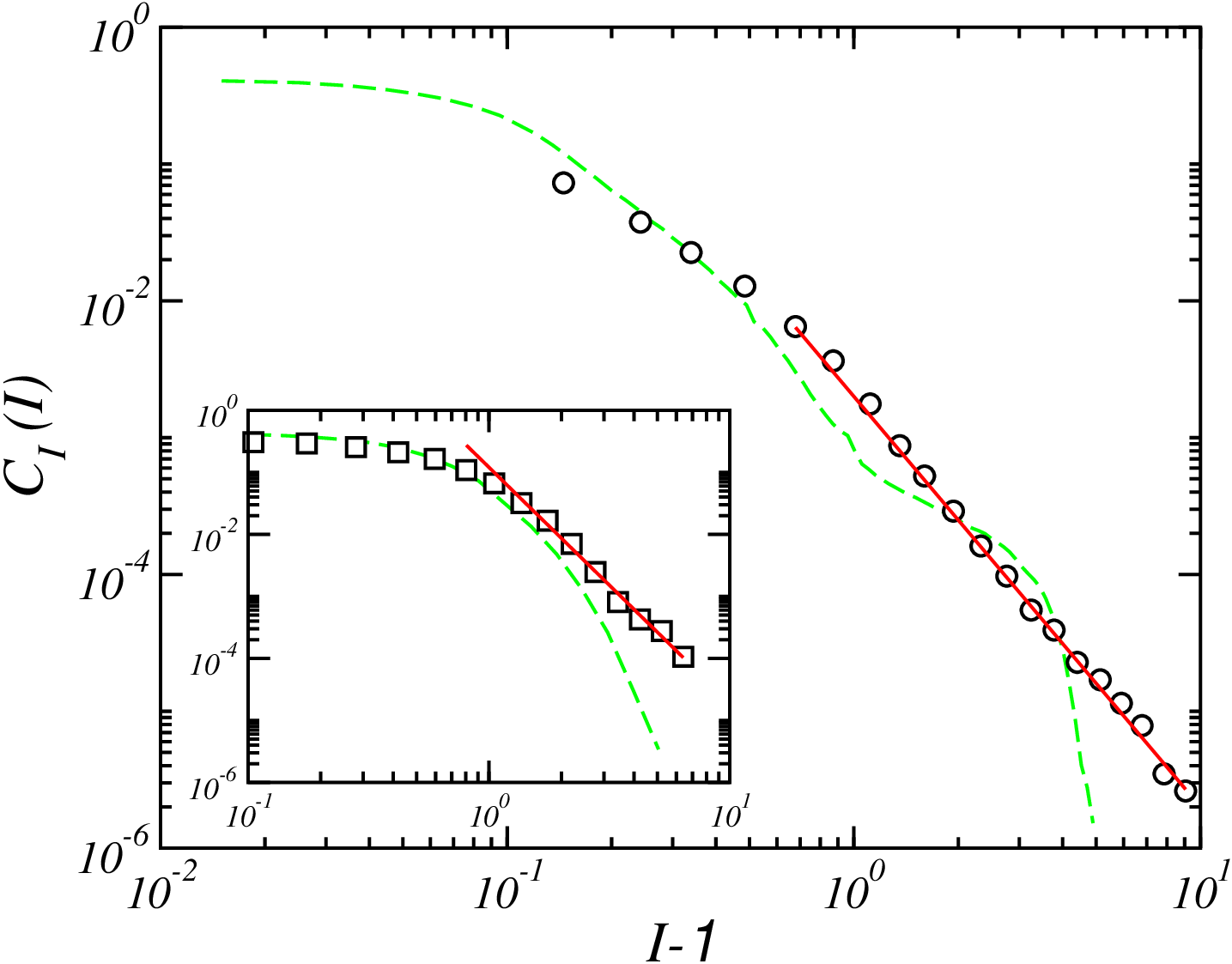}
\caption{Cumulative distribution for the inertia of the actors in the movie
database. In the inset, the same distribution for database of publication on
biosensors. The straight red (continuous)
lines are only indicative and correspond to power laws with exponents
$-3$ in the main plot and  $-3.8$ in the inset. The green (dashed) curves are simulation
results.}
\label{pi}
\end{figure}

One of the easier to detect effects pointing to the necessity of the
use of weighted networks to describe collaboration graphs
is the behavior of $\langle k \rangle_q$ as illustrated in Fig. \ref{kq}. The 
meaning of this quantity 
is the average total number of different
partners that an actor with experience $q$ has had along his/her
professional career. In an idealized situation where the actors did
not show any tendency to repeat collaborators, this magnitude should
follow a linear growth with $q$. $\langle k \rangle_q$ would
approach $(\langle n \rangle -1) \, q$, or even $(\langle n \rangle
-1) \, q/\langle w \rangle$ if we admit the existence of a sharp
$P_w(w)$ distribution centered around $\langle w\rangle$. However,
the empirically measured $\langle k \rangle_q$ functions do not show 
linear growth in any of our 
networks. These functions are better fitted by power laws with exponents 
in the range from
$0.5$ to $0.8$, with reservations due to  the short $q$ range 
in some cases, than by straight lines. The reason behind this peculiar 
behavior of $\langle k \rangle_q$ is
the nontrivial structure of $P_w(w)$. Instead of being a Gaussian or
some other smooth distribution centered around $\langle w
\rangle$, $P_w(w)$ falls in a power law-like way for high $w$ as may
be seen in Fig. \ref{pw}. Actually, for most of our networks
$P_w(w)$ adjusts better to a power law than the distributions
$P_q(q)$ or $P_k(k)$. Figure \ref{pw}, instead of plotting
$P_w(w)$, plots the cumulative distributions $C_w(w) = \int_w^\infty
dw'\, P_w(w')$ and $C_s(s) = \int_s^\infty ds'\, P_s(s')$ for the
movie network. Note that if $P_w(w) \sim w^{-\delta}$, $C_w(w) \sim
w^{-\delta+1}$. The estimated values for the exponent $\delta$
obtained from the different databases are listed in Table I. Typically, the value of $\delta$ is high, between $3$ and $6$.
Nevertheless, the distribution $P_w(w)$ seems in general to be incompatible
with any faster decay as for instance an exponential tail.

\begin{figure}
\epsfxsize=80mm \epsffile{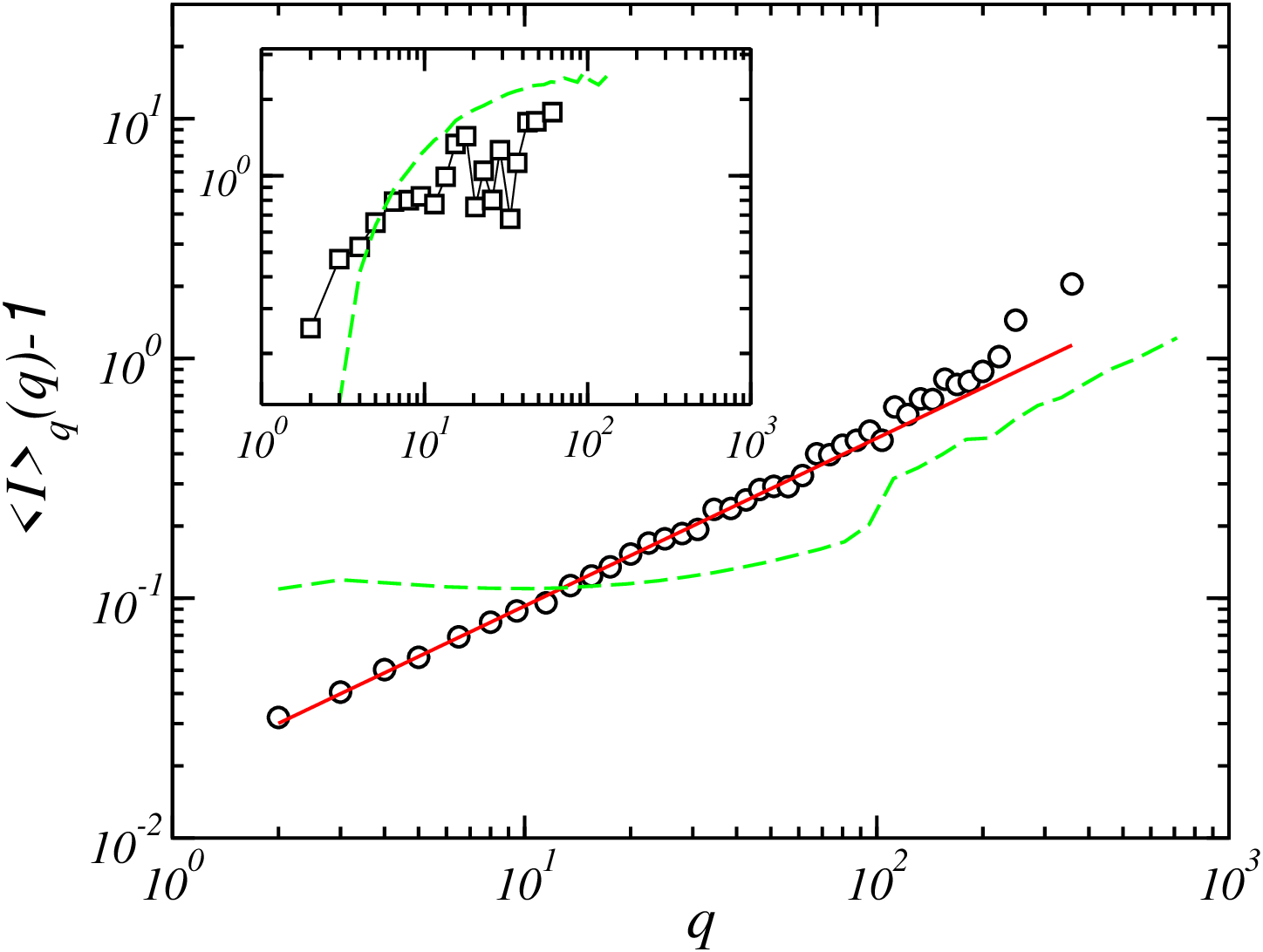}
\caption{Average inertia of the actor with a certain experience $q$ as a function
of $q$. The main plot is for the movie database and the inset for the
scientific papers on superstrings. The straight red (continuous) line in the main plot is a
power laws with exponent $0.7$. The green (dashed) curves correspond to numerical
simulation of the theoretical model.}
\label{iq}
\end{figure}

As mentioned above, for weighted bipartite networks in the absence
of correlations there are only three independent distributions. This
means that the functional shape of $P_\mathcal{I}(\mathcal{I})$
should be a consequence of the shapes of other distributions like
$P_n(n)$, $P_q(q)$ and $P_w(w)$. In Figure \ref{pi}, the cumulative
distribution $C_\mathcal{I} = \int_\mathcal{I}^\infty d\mathcal{I}'
\, P_\mathcal{I}(\mathcal{I}')$ is displayed for the movie and the
scientific publication on biosensor databases. In this figure, we
have plotted the cumulative inertia distribution versus
$\mathcal{I}-1$. The distribution represented in this way shows an
initial offset followed by a relatively long tail for high values of
$\mathcal{I}$. It is hard to ascertain whether the asymptotic
behavior of $P_\mathcal{I}(\mathcal{I})$ continues or not because of
the limited range of values of $\mathcal{I}$. The inertia is defined
as a ratio between the strength and the degree and hence a certain
value of $\mathcal{I}$ means that the strength is actually
$\mathcal{I}$ times bigger than the degree. Therefore, the values
that the inertia can attain are conditioned very strongly by the
network size. In the rest of the networks, the main trend presented
in Fig. \ref{pi} repeats even though in some cases the network size
is too small to reliably determine the statistical significance of the
results.

Another interesting feature of inertia is its dependence on the age of
the actors. One would expect the older actors to be more
conservative but is it really the case? We cannot study this issue
directly because our databases do not contain the age of the
actors. The best we can do is to associate the age with the
experience $q$. Hence, we have depicted in Fig. \ref{iq} the average
inertia of all the actors with a certain experience $q$ as a
function of $q$ for the databases on movies and on scientific
publications on superstrings. In these two cases, contrasting
behavior may be observed. The actors in the movie database, similar
to the authors of some of the scientific publication databases
(e.g., atrial ablation, botox), become increasingly conservative as
they acquire experience, although the increase in $\langle
\mathcal{I} \rangle_q$ is not linear with $q$. While in the case of
the superstring community, as well as in other scientific
specialties (biosensors, condmat, etc) there is a saturation in the
value that $\langle \mathcal{I} \rangle_q$ can attain with $q$.

\begin{figure}
\epsfxsize=80mm \epsffile{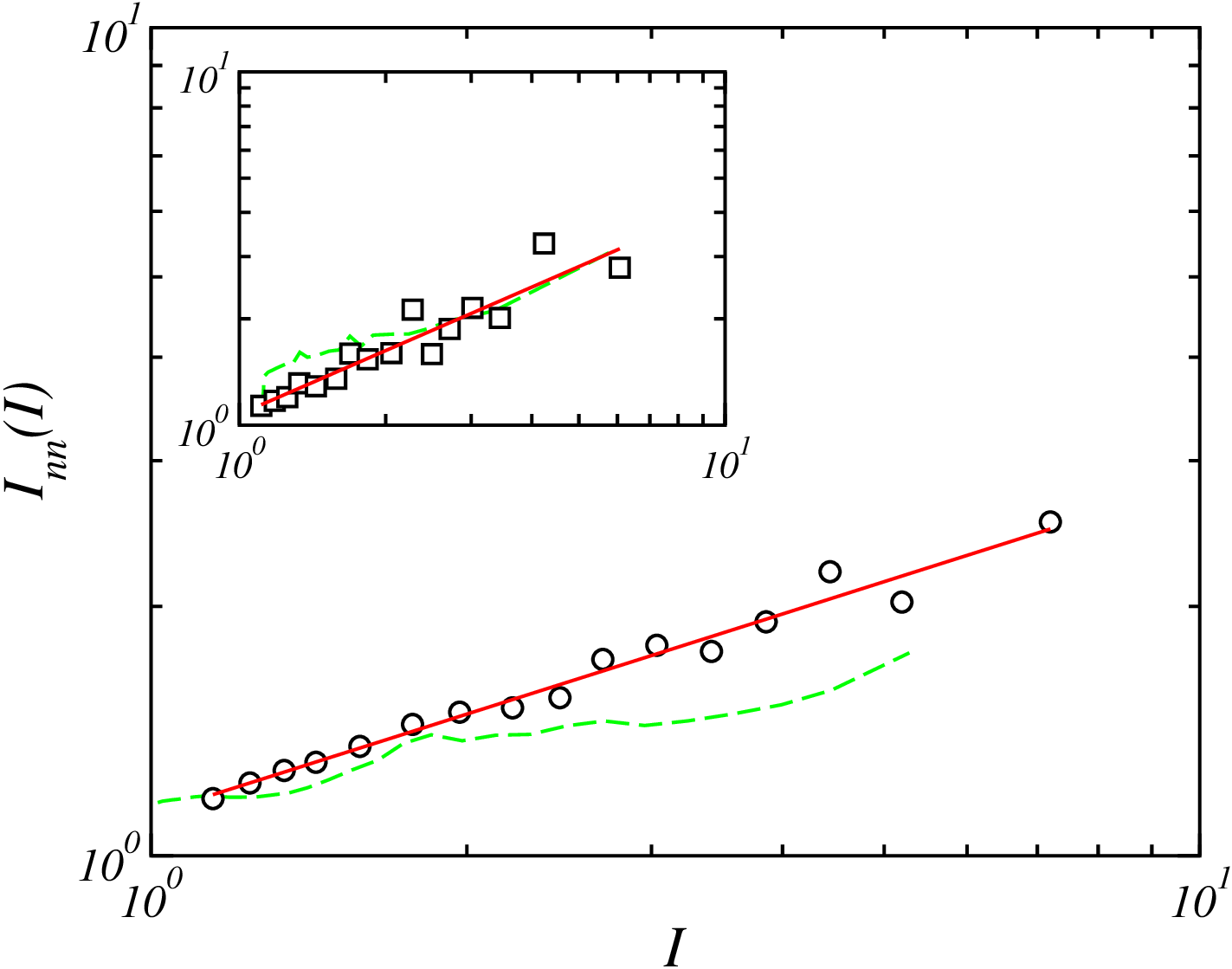}
\caption{Mean inertia of the nearest neighbors as a function of the own
inertia. The main plot correspond to the data of the IMDB movie database and
the inset to the publications on info science. The two straight red (continuous)
lines have
slopes $0.4$ in the main plot and $0.6$ in the inset. The green (dashed) curves are
simulation results.}
\label{inni}
\end{figure}

The last aspect that we shall contemplate about the inertia is the
correlations between the inertia of neighboring actors. Do
conservative actors like to collaborate with conservative
counterparts? The answer to this question is displayed in Fig.
\ref{inni}. There the average inertia of the nearest neighbors is
depicted as a function of the own inertia for the movie database and
for the publication on info science. In both cases, as in all the
other databases studied, it is clear that there is a positive correlation
of an author's inertia to the inertia of the author's collaborators.
 In some of the networks the growth appears linear, but in some 
 others, e.g., the movie database, the curves are
better fitted by power laws with exponents as low as $0.4$.

\section{Theoretical model}
\label{models}

Several models have been proposed to mimic the development of
collaboration networks \cite{newman01,guillaume03,fan04,borner05}. In this 
work, we shall focus on one model that combines a certain simplicity in
 the rules with acceptable results for the
topology of the network, including properties such as the
correlations or the size of the giant component. It was proposed and studied
in Refs. \cite{ramasco04,morris04,guimera05,peltomaki05}. In this 
section,
we will address the question whether this model is able or not to reproduce 
the empirically observed behavior of actors social inertia. The
rules of the model are as follows:
\begin{enumerate}

\item At each step a new collaboration of size $n$ is introduced in the system.
$n$ may be a fixed external parameter or may be obtained as a random
variable from a distribution $P_n(n)$. We use this latter option
with an exponential decaying $P_n(n)$ that is the closest functional
form to the empirical observed collaboration size distribution. This
means that the parameter $\langle n \rangle$ has to be externally
provided.

\item Out of the $n$ collaborators, $m$ are newcomers without previous
experience. Again $m$ may be a fixed external number or it may be
derived from a random distribution. We have checked both
possibilities, taking always into account the obvious constraint $m
\leq n$.

\item The remaining $n-m$ actors are chosen from the pool of experienced
individuals in the following way:

     \begin{enumerate}

     \item with probability $p$, one of the experienced actors already
     in the present collaboration (if there is some) is
     randomly selected and one of his/her
     partners from previous collaborations is chosen to participate in the new
     collaboration with a
     probability proportional to the number of times they have worked
     together.

     \item with probability $1-p$, an experienced actor is selected with
     a probability proportional to his/her experience $q$.

     \end{enumerate}

\item After the collaboration is complete, each actor updates his/her
experience $q_i' \to q_i+1$. The actors can then become inactive, ineligible for the
previous rule, if their experience is higher than an externally introduced
threshold, $q > Qc$, with a probability $1/\tau$.

\end{enumerate}

\begin{table*}
\label{mod-tab} \caption{Global parameters of the simulated
networks.}
\begin{ruledtabular}
\begin{tabular}{ccccccccccccc}
Simulated network& $m$& $\langle n \rangle$& $p$ & $Q_c$ & $\tau$&
$N_{ai}/N_{a} (\%)$& $\langle q \rangle$& $\langle k \rangle$&
$\langle \mathcal{R} \rangle$& $\langle \mathcal{I} \rangle$&
$\langle C \rangle$&
$\langle C^w \rangle$\\
\hline

movies& $3.0$& $11.5$& $0.79$& $100$& $150$ & $2.5$& $5.3$& $89.1$&
$0.04$& $1.05$& $0.76$&
$0.38$\\

biosensors& $1.87$& $3.89$& $1.$& $30$& $18$ & $13.6$& $3.4$&
$10.7$& $0.014$& $1.24$& $0.81$&
$0.41$\\

condmat& $0.76$& $2.66$& $0.87$& $15$& $11.5$ & $21.2$& $7.0$&
$10.0$& $0.024$& $1.44$& $0.63$&
$0.34$\\

info science& $0.66$& $1.38$& $0.86$& $50$& $23$ & $0.75$& $4.82$&
$0.46$& $0.02$& $1.44$& $0.71$&
$0.36$\\

superstrings& $0.57$& $2.04$& $1.$& $20$& $15$ & $32.0$& $8.7$&
$4.34$& $0.36$& $1.97$& $0.85$&
$0.46$\\

\end{tabular}
\end{ruledtabular}
\end{table*}

The last rule is introduced to account for the limited professional life
time of actors. This fact has also an important impact on
the network correlations as was shown in \cite{ramasco04}, only 
contemporaneous active actors can carry out a work together. The model has hence five
external parameters apart form the network size. Some like $\langle n \rangle$
and $m = N_a/N_c$ are easy to estimate from empirical data. To approach $p$,
we must consider the two sources for new edges during the growth 
process: the newcomers and the old actors added by rule $3b$. 
The probability $p$ can be then approximated from the empirical
values of $R_{ks}$, and the moments $\langle n \rangle$ and $\langle
n^2 \rangle$ of the collaboration size distribution $P_n(n)$.
Finally, $Q_c$ and $\tau$ are chosen according to the trends
observed in the empirical $P_q(q)$ distributions. The values of
these parameters used in the simulations, together with the results
for the main global magnitudes characterizing the topology of the
networks, are displayed in Table II. We have selected the parameters
of the largest empirical networks for the simulation. In each case, the 
number of collaborations simulated is $N_c = 10^5$.

As may be seen in Table II, the model reproduces relatively well the
global parameters for the first three networks (movies, biosensors
and condmat) but fails to do so for the other two networks (info
science and superstrings). The reason might be the
different degree of heterogeneity of the databases. IMDB and condmat
comprehend the output of several independent communities: the IMDB
includes a variety of movie genders that can be assumed to be
starred by separate non-overlapping groups of actors, and the papers
submitted to condmat deal with a range of topics from
experimental and theoretical solid state physics to statistical
physics produced by diversified scientific communities.
These heterogeneous networks are more suitable to be modeled
just with a simple set of general rules. On the hand, publications
on superstrings and on info science should correspond to more
homogeneous scientific communities where other factors such as citations 
can have an important impact.

More detailed results from the simulations are represented in
Figures \ref{kq} to \ref{inni}, green (dashed) curves. The model
fails in all cases to mimic the nonlinear dependence of 
$\langle k \rangle_q$ with $q$. Despite of that, the
cumulative distributions $C_w(w)$, $C_s(s)$ and
$C_\mathcal{I}(\mathcal{I})$ (Figs \ref{pw} and \ref{pi}) are
qualitatively well reproduced by the model with perhaps some minor
problems at the very end of the tails due to the aging mechanism. Regarding 
the correlations, the predicted $\langle \mathcal{I}
\rangle_q$ adjust well to the superstring data (see inset
of Figure \ref{iq}) reproducing even the saturation observed. It does not 
match so well though the behavior of the
IMDB movie network. Finally, the inertia-inertia
correlation trend, more conservative actors tend to collaborate with
conservative counterparts, is also observed in the model. Although, the 
agreement with the empirical networks is only qualitative.

\section{Conclusions}
\label{conclusion}

In summary, we have studied collaboration networks of several
disciplines using weighted networks to represent the one-mode
projections of the full bipartite network. This representation
allows us to define social inertia, as the ratio between the number
of collaborators and the total number of partnerships. This new metric can be in
general 
defined in all weighted graphs. It has a very special meaning though in 
the specific case of social networks: it quantitatively measures the 
tendency of the actors to repeat the
same collaborators. We have shown that the inertia of the actors in
empirical networks displays features characteristic of complex
systems. The distribution of $\mathcal{I}$ is long tailed and its
dependence with the experience or the correlations between the
inertia of coauthors are far from trivial. We have found that the
inertia generally grows with the experience though it saturates for
some networks. At the same time, we have also shown that
conservative actors have a strong tendency (that can be quantified)
to collaborate with conservative actors. This is, we hope, another
effort towards a more quantitative Sociology taking advantage of the
developments of other branches of Science such as Statistical
Physics and Graph Theory.

We have also studied the predictions of a theoretical growth model
for collaboration networks. This model is the simplest that is able
to reproduce some evolved topological properties of empirical
networks such as the degree-degree correlations. The results of the
simulation are in qualitative agreement with the real networks
observations. However, in the search for a more quantitative insight
of the network growth process there are still several open issues.
One is the connection of the co-authorship network development with
other aspects of the system as, for example citation networks that has been
discussed in recent publications, and its quantification. Another is
a more detailed study about the influence that some other author
factors as age, scientific structures such as research groups, projects or even 
big research facilities may have on the topology of the collaboration
network.

\begin{acknowledgments}
The authors gratefully thank Stefan Boettcher and Sergei Dorogovtsev for useful discussion and
comments. Partial funding from the NSF under grant 0312510 was
received. Thanks to Chaomei Chen, Katy Borner, Soren Paris, and 
M.E.J Newman for supplying
the superstrings, info viz, scientometrics, and condmat data, respectively.
All other journal paper databases were gathered
by S.A. Morris.
\end{acknowledgments}


\begin{thebibliography}{10}

\bibitem{barabasi02}
R. Albert and A.-L. Barab{\'a}si, Rev. Mod. Phys. {\bf 74},  47  (2002).

\bibitem{sergei03}
S.N. Dorogovtsev and J.F.F. Mendes, {\em Evolution of networks: From
  Biological Nets to the Internet and WWW} (Oxford University Press,
  Oxford, 2003).

\bibitem{romu04}
R. Pastor-Satorras and A. Vespignani, {\em Evolution and structure of the
  Internet: A statistical physics approach} (Cambridge University Press,
  Cambridge, 2004).

\bibitem{newman03}
M.E.J. Newman, SIAM Review {\bf 45},  167  (2003).

\bibitem{barabasi99}
A.-L. Barab\'asi and R. Albert, Science {\bf 286}, 509 (1999).

\bibitem{albert99}
R. Albert, H. Jeong, and A.-L. Barab{\'a}si, Nature {\bf 401},  130  (1999).

\bibitem{romu01}
R. Pastor-Satorras, A. V{\'a}zquez, and A. Vespignani,
Phys. Rev. Lett. {\bf 87},  258701  (2001).

\bibitem{jeong00}
H. Jeong, B. Tombor, R. Albert, Z.N. Oltvai, and A.-L. Barab\'asi,
Nature {\bf 407}, 651 (2000).

\bibitem{schwikowski00}
B. Schwikowski, P. Uetz, and S. Fields, Nat. Biotech. {\bf 18}, 1257 (2000).

\bibitem{wuchty03}
S. Wuchty, Z.N. Oltvai, and A.-L. Barab\'asi, Nat. Genet. {\bf 35}, 176 (2003).

\bibitem{newman01}
M.E.J. Newman, S.H. Strogatz, and D.J. Watts, Phys. Rev. E {\bf 64}, 026118
(2001); Proc. Natl. Acad. Sci. USA {\bf 99}, 2566 (2002).

\bibitem{newman03b}
M.E.J. Newman and J. Park, Phys. Rev. E {\bf 68}, 036122 (2003).

\bibitem{newman01b}
M.E.J. Newman, Proc. Natl. Acad. Sci. USA {\bf 98}, 404 (2001); Phys. Rev. E
{\bf 64}, 016131 and 016132 (2001).

\bibitem{note0} Note that not even the representation of the one--mode 
projected network
as a weighted graph contains the same degree of information as the full
bipartite network. In order to reconstruct the full bipartite graph from the
one--more projection, it would be necessary to know exactly to which movie every
unit of the link weigth corresponds. 

\bibitem{yook01}
S.H. Yook, H. Jeong, A.-L. Barab\'asi, and Y. Tu, Phys. Rev. Lett. {\bf 86}, 5835
(2001).

\bibitem{barrat04}
A. Barrat, M. Barth\'elemy, R. Pastor-Satorras, and A. Vespignani, Proc. Natl.
Acad. Sci. USA {\bf 101}, 3747 (2004).

\bibitem{barthelemy04}
M. Barth\'elemy, A. Barrat, R. Pastor-Satorras, and A. Vespignani, Physica A
{\bf 346}, 34 (2004).

\bibitem{fan04}
Y. Fan, M. Li, J. Chen, L. Gao, Z. Di, and J. Wu, Int. J. Mod. Phys. B {\bf
18}, 2505 (2004).

\bibitem{borner05}
K. B\"orner, L. Dall'Asta, W. Ke, and A. Vespignani, e-print cond-mat/0502147.

\bibitem{newman02}
M.E.J. Newman, Phys. Rev. Lett. {\bf 89}, 208701 (2002).

\bibitem{newman03c}
M.E.J. Newman, Phys. Rev. E {\bf 67}, 026126 (2003).

\bibitem{note1}
Data available at
\texttt{http://www.nd.edu/$\sim$networks/dat\-ab\-a\-s\-e/index.html}.

\bibitem{note2}
Database located at
\texttt{http://arxiv.org/ar\-chi\-ve/cond-mat}.

\bibitem{nets-Stev} Databases available at 
\texttt{http://samorris.ceat.okstate.\-edu/web/matrices/ap\_p/default.htm}.

\bibitem{atrial-dis} S.A. Morris, M.L. Goldstein and C.F. DeYong, sumitted work.

\bibitem{botox} C.M. Chen, and S.A. Morris, Proc. IEEE Symposium on 
Information Visualization, Seattle, Washington, 67 (2003).

\bibitem{complex} M.L. Goldstein, S.A. Morris, and G. Yen,
Eur. Phys. J. B {\bf 41}, 255 (2004).

\bibitem{superstring} C. Chen and J. Kuljis, Journal of the American Society 
for Information Science and Technology {\bf 54}, 435 (2003).

\bibitem{guillaume03} J.-L. Guillaume and M. Latapy, e-print cond-mat/0307095.

\bibitem{ramasco04} J.J. Ramasco, S.N. Dorogovtsev and R. Pastor-Satorras, Phys.
Rev. E {\bf 70}, 036106 (2004).

\bibitem{morris04} M.L. Goldstein, S.A. Morris and G.G. Yen,
Phys. Rev. E {\bf 71}, 026108 (2005); S.A. Morris, e-print cond-mat/0501386.

\bibitem{morris05} S.A. Morris and G.G. Yen, e-print
physics/0503061.

\bibitem{guimera05} R. Guimer\`a, B. Uzzi, J. Spiro and L.A.N. Amaral, Science
{\bf 308}, 697 (2005).

\bibitem{peltomaki05} M. Peltom\"aki and M. Alava, e-print cond-mat/0508027.

\end{thebibliography}
\end{document}